\documentstyle{amsppt}
\newcount\refcount
\advance\refcount 1
\def\newref#1{\xdef#1{\the\refcount}\advance\refcount 1}
\newref\cosets
\newref\crssII
\newref\gottesman
\newref\knilllaflamme
\newref\invariants
\newref\shadow
\newref\nonadditive
\newref\shorlaflamme

\def\Tr{\operatorname{Tr}}

\def\tensor{\otimes}
\def\omegabar{\overline{\omega}}
\topmatter
\title Quantum codes of minimum distance two \endtitle
\author Eric M. Rains\endauthor
\affil AT\&T Research \endaffil
\address AT\&T Research, Room 2D-147, 600 Mountain Ave.
         Murray Hill, NJ 07974, USA \endaddress
\email rains\@research.att.com \endemail
\date April 4, 1997\enddate
\abstract
It is reasonable to expect the theory of quantum codes to be simplified in
the case of codes of minimum distance 2; thus, it makes sense to examine
such codes in the hopes that techniques that prove effective there will
generalize.  With this in mind, we present a number of results on codes of
minimum distance 2.  We first compute the linear programming bound on the
dimension of such a code, then show that this bound can only be attained
when the code either is of even length, or is of length 3 or 5.  We next
consider questions of uniqueness, showing that the optimal code of length 2
or 4 is unique (implying that the well-known one-qubit-in-five single-error
correcting code is unique), and presenting nonadditive optimal codes of all
greater even lengths.  Finally, we compute the full automorphism group of
the more important distance 2 codes, allowing us to determine the full
automorphism group of any $GF(4)$-linear code.

\endabstract
\endtopmatter
In classical coding theory, the simplest nontrivial codes are the
codes of minimum distance 2, and their duals, repetition codes.  Here,
binary codes of minimum distance 2 are unique (consisting of all even
weight vectors), linear, and essentially trivial to analyze.  In the
quantum setting, as we shall see, the situation is not nearly so nice.
The purpose of the present work is to explore the structure of quantum
codes of minimum distance 2, both because the theory is likely to be
simpler in that case, and because many codes of interest (e.g.,
$GF(4)$-linear codes) are built out of distance 2 codes.

The first natural question to ask about distance 2 codes is: How good
can they be?  That is, how large can the dimension of a distance 2
code be, given its length?  In the case of even length $2m$, this is easy;
the quantum Singleton bound (\cite\knilllaflamme) states that the dimension
can be at most $4^{m-1}$, while the additive code
$$
\vbox{\halign{&$#$\cr
1      & 1      & 1      & \ldots & 1\cr
\omega & \omega & \omega & \ldots & \omega\cr
}}\tag 1
$$
attains this bound.  For odd length, the situation is not as simple as this;
however, one can use linear programming to give a bound.  The resulting
bound could, in principle, be met whenever the length is of the form
$2^m+1$ (and is, indeed, met, when $m=2$ (\cite\nonadditive));
however, as we shall see, this does not happen.  We also give a construction
which, given a pure $((n,K,2))$, produces a pure $((n+2,4K,2))$, thus
giving a lower bound of $3\cdot 2^{n-4}$ on the optimal dimension for $n$ odd.

Since the code \thetag{1}, which we shall refer to as ``the
$[[2n,2(n-1),2]]$'' in the sequel, has a particularly nice structure, with
a very large symmetry group, it is reasonable to conjecture that {\it any}
optimal distance 2 code of even length is of that form.  It is easy to see
that this is, indeed, the case for length 2; on the other hand, we
construct a nonadditive optimal code of length 6, which allows the
construction of nonadditive optimal codes of all larger even lengths.  This
leaves only length 4 open; by considering the quartic invariants of such a
code (\cite\invariants), we show that the optimal length 4 code is unique
as well.  As the well-known $[[5,1,3]]$ and its associated self-dual
$[[6,0,4]]$ can be built up from $((4,4,2))$s, we can show that those codes
are unique as well.

Finally, we consider the question of the full automorphism group of the
codes \thetag{1}; that is, automorphisms that do not necessarily lie in the
Clifford group.  This question is important, for instance, because
automorphisms of a code induce fault-tolerant operations (\cite\gottesman).
Here, we find that such nonadditive automorphisms can occur only for length
2.  Since $GF(4)$-linear codes are built out of $[[2n,2(n-1),2]]$s, we
find, for instance, that any equivalence between $GF(4)$-linear codes
(subject to certain trivial restrictions) must lie in the Clifford group.

We recall the notation that a $[[n,k,d]]$ refers to an additive code
encoding $k$ qubits in $n$ qubits, with minimum distance $d$; a $((n,K,d))$
refers to a general code encoding $K$ states in $n$ qubits with minimum
distance $d$.  Also, we will say that two codes are locally equivalent if
there is an element of $PSU(2)^{\tensor n}$ mapping one into the other; we
will say they are globally equivalent, or just equivalent, if it is
possible to permute the qubits of one in such a way as to make it locally
equivalent to the other.  A code will be said to be nonadditive if it is
not equivalent to any additive code.

\head Bounds \endhead

We first recall the following fact:

\proclaim{Theorem 1}
Let $Q$ be a $((2m,K,2))$ for some $m$ and $K$.  Then $K\le 4^{m-1}$.
On the other hand, for all $m\ge 1$, there exists an $((2m,4^{m-1},2))$.
\endproclaim

\demo{Proof}
The quantum Singleton bound (\cite\knilllaflamme), states that for any
$((n,K,d))$, we must have
$$
K\le 2^{n-2(d-1)}.
$$
Applying this for $d=2$, we get the stated bound.

As remarked in the introduction, existence follows from consideration of
the code \thetag{1}.
\qed\enddemo

For odd length, the linear programming bound (\cite\shadow,
\cite\shorlaflamme; note that the Singleton bound is a special case) is
more complicated, but still feasible to compute:

\proclaim{Theorem 2}
Let $Q$ be a $((2m+1,K,2))$ for some $m$ and $K$.  Then
$$
K\le 4^{m-1} (2-{1\over m}).\tag 2
$$
\endproclaim

\demo{Proof}
We consider the coefficients $B_0$, $B_1$, and $S_0$ of the dual and
shadow enumerators of $Q$.  These can be expressed in terms of the
weight enumerator of $Q$ as follows:
$$
\align
B_0&=2^{-n} \sum_{0\le i\le n} A_i,\\
B_1&=2^{-n} \sum_{0\le i\le n} (3n-i) A_i,\\
S_0&=2^{-n} \sum_{0\le i\le n} (-1)^i A_i,
\endalign
$$
where $n=2m+1$.  Eliminating $A_n$ and $A_{n-1}$, we get:
$$
(n-2) B_0+B_1-2 S_0
=
2^{-n} \sum_{0\le i\le m}
4(n-2i-1) (A_{2i}+A_{2i+1}).
$$
Now, since $Q$ has minimum distance 2, it follows that
$K B_0=A_0$, and $K B_1=A_1$.  Substituting in, we find:
$$
\multline
(K^{-1}(n-2)-2^{-n}(4n-4)) A_0+
(K^{-1}-2^{-n}(4n-4)) A_1\\
=
2 S_0+
2^{-n} \sum_{1\le i<m}
4(n-2i-1) (A_{2i}+A_{2i+1}).
\endmultline
$$
The coefficients on the right-hand-side are positive, while the coefficients
on the left-hand-side are negative whenever $K>4^{m-1} (2-{1\over m})$.
This contradicts the fact that $A_0>0$, while all other $A_i$ and $S_0$
are nonnegative.
\qed\enddemo

It is straightforward to verifiy that the enumerator
$$
(4^{m-1}(2-{1\over m}))^2 (x^n+{n\over n-2} x y^{n-1}+{2n-2\over n-2} y^n)
$$
is feasible for $K$ meeting the bound \thetag{2}; thus this bound agrees
with the full linear programming bound.

The bound \thetag{2} is integral precisely when $m$ is a power of 2; thus,
it is natural to conjecture that it can be met then.  For $n=3$, the bound
is $K\le 1$, met, for instance, by the self-dual additive code:
$$
\vbox{\halign{&$#$\cr
1      & 1      & 0      \cr
0      & 1      & 1      \cr
\omega & \omega & \omega \cr
}}
$$
For $n=5$, the bound is $K\le 6$, which is attained by the $((5,6,2))$
given in \cite\nonadditive.  Thus, it is somewhat surprising that these
are the only cases in which the bound can be met:

\proclaim{Theorem 3}
For no $i\ge 3$ is there a $((2^i+1,2^{2^i-1}-2^{2^i-i-1},2))$.
\endproclaim

\demo{Proof}
Suppose $Q$ were such a code.  From the proof of theorem 2, we can conclude
that $S_0=0$, and $A_i=0$ for $1\le i\le n-2$.  Solving the equations
$A_0=K^2$, $B_0=K$, and $K B_1=A_1$, we find that the weight enumerator of
$Q$ must be
$$
(2^{2^i-1}-2^{2^i-i-1})^2 (x^{2^i+1}+{2^i+1\over 2^i-1} x y^{2^i}+
{2^{i+1}\over 2^i-1} y^{2^i+1}).
$$
Now, since $Q$ is pure of minimum distance 2, if we trace away one qubit
of $Q$, the resulting operator must be $P_{Q'}/2$, where $Q'$ is a
$((2^i,2^{2^i}-2^{2^i-i},2))$ with weight enumerator
$$
(2^{2^i}-2^{2^i-i})^2 (x^{2^i}+{1\over 2^i-1} y^{2^i});
$$
the coefficient of $y^{2^i}$ follows from the fact that $B_0(Q')=K'$.
Consider the orthogonal complement of $Q'$.  We can compute its
weight enumerator, by noting that orthogonal complementation leaves
each $A_i$ unchanged, except that it changes $A_0$ from $K^2$ to $(2^n-K)^2$.
Thus $Q'$ has weight enumerator
$$
(2^{2^i-i})^2 (
x^{2^i} +
(2^i-1) y^{2^i}),
$$
and dual enumerator
$$
2^{2^i-i} (
x^{2^i}+
(4-2^i) x^{2^i-1} y+\ldots).
$$
For $i>2$, $4-2^i$ is negative, giving a contradiction.
\qed\enddemo

Remark.  This result is quite fragile; in particular, it depends strongly
on the fact that the code $Q$ must be pure.  Consequently, the above
argument cannot be used to strengthen the bound \thetag{2} {\it except} to say
that it cannot be met.  It should be possible to give a different proof
of theorem 3 that works for slightly suboptimal enumerators as well.

Remark. For $i=2$, the orthogonal complement of the code $Q'$ has weight
enumerator
$$
16 x^4+48 y^4,
$$
which is the weight enumerator of a $((4,4,2))$; we will use this fact
in the sequel.

The best lower bound we have been able to prove is the following:

\proclaim{Theorem 4}
For all $m\ge 2$, there exists a pure $((2m+1,3\cdot 2^{2m-3},2))$.
\endproclaim

\demo{Proof}
This was shown for $m=2$ in \cite\nonadditive.  For $m>2$, we will need
the following lemma:

\proclaim{Lemma 5}
If there exists a pure $((n,K,2))$, then there exists a pure
$((n+2,4K,2))$.
\endproclaim

\demo{Proof}
Let $Q$ be a pure $((n,K,2))$, and let $v_2$ be the state
$|00\rangle+|11\rangle$ on two qubits.  Then the new code $Q'$ is the span of:
$$
\{
Q\tensor v_2,
(X_1\tensor X_1) (Q\tensor v_2),
(Y_1\tensor Y_1) (Q\tensor v_2),
(Z_1\tensor Z_1) (Q\tensor v_2)\},
$$
where $X_1$ is the operator that acts as $\sigma_x$ on the first qubit and
as the identity on the remaining qubits, and similarly for $Y_1$ and $Z_1$.

Examination of the appended pair of qubits tells us that $Q'$ has four times
the dimension of $Q$; it remains only to show that $Q'$ is pure of
minimum distance 2.  In other words, we need to show that any single-qubit
error takes $Q'$ to an orthogonal code; this is straightforward to verify.
\qed\enddemo

\noindent The theorem follows by induction.
\qed\enddemo

Lemma 5 suggests the following natural question: What is the value of
$$
\kappa_2=\lim_{m\to\infty} 4^{1-m}K_0(2m+1),
$$
where $K_0(2m+1)$ is the maximum dimension of a pure $(2m+1,K,2)$?  By
the lemma, this sequence is nondecreasing, so the limit exists, and is
bounded between ${3\over 2}$ (by theorem 4), and 2 (by theorem 2).

\head Uniqueness results\endhead

It is natural to wonder whether the codes $(1)$ are necessarily the {\it only}
optimal distance 2 codes of even length.  For length 2, the answer is an
easy ``Yes''; indeed, one can show the following stronger result:

\proclaim{Lemma 6}
Let $w_1$, $w_2$, $w_3$, and $w_4$ be an orthonormal basis of
${\Bbb C}^2\tensor {\Bbb C}^2$ consisting of $((2,1,2))$s.  Then
there exists elements $U_1$ and $U_2$ of $SU(2)$ such that
$$
\align
(U_1\tensor U_2) w_1&\propto (|00\rangle+|11\rangle)/\sqrt{2}\\
(U_1\tensor U_2) w_2&\propto (|01\rangle+|10\rangle)/\sqrt{2}\\
(U_1\tensor U_2) w_3&\propto -i(|01\rangle-|10\rangle)/\sqrt{2}\\
(U_1\tensor U_2) w_4&\propto (|00\rangle-|11\rangle)/\sqrt{2}.
\endalign
$$
\endproclaim

\demo{Proof}
Let $w=a|00\rangle+b|01\rangle+c|10\rangle+d|11\rangle$ have norm 1.
For $w$ to have distance 2, we must have
$$
\Tr_1(w w^\dagger)=\Tr_2(w w^\dagger)=I/2.
$$
This gives the conditions:
$$
\align
|a|^2+|c|^2&=1/2\\
|b|^2+|d|^2&=1/2\\
a \overline{b} +c \overline{d}&=0.
\endalign
$$
In other words, the matrix
$$
M(w)=\sqrt{2}\pmatrix a&b\\c&d\endpmatrix
$$
must be unitary.  Thus the theorem is equivalent to the statement
that, for any 2 by 2 unitary matrices $M_1$, $M_2$, $M_3$, and
$M_4$, orthonormal under the inner product ${1\over 2}\Tr(A B^\dagger)$,
there exist unitary matrices $U_1$ and $U_2$ such that
$$
U_1 M_1 U^t_2\propto I,\quad
U_1 M_2 U^t_2\propto \sigma_x,\quad
U_1 M_3 U^t_2\propto \sigma_y,\quad
U_1 M_4 U^t_2\propto \sigma_z.
$$
To satisfy the first equation, we may take $U_2 = \overline{U_1 M_1}$;
we may therefore assume $M_1=1$, without loss of generality.  Then, up to
phase, we have $M_2^2=M_3^2=M_4^2=1$.  It follows that each of $M_2$,
$M_3$, and $M_4$ can be written as real linear combinations of $\sigma_x$,
$\sigma_y$, and $\sigma_y$; this determines an orthonormal basis of ${\Bbb
R}^3$. Conjugation by $SU(2)$ acts as $SO(3)$ on ${\Bbb R}^3$;
consequently, we may take the given basis to the standard basis.  The
resulting transformation gives us unitary matrices as desired.
\qed\enddemo

Thus, in particular, any $((2,1,2))$ is locally equivalent to the $[[2,0,2]]$.

For $n\ge 6$, on the other hand, the answer is ``No'':

\proclaim{Theorem 7}
For all even $m\ge 3$, there exists a nonadditive $((2m,4^{m-1},2))$.
\endproclaim

\demo{Proof}
In light of lemma 5, it suffices to prove the result for $m=3$, which we
may do by explicit construction.  Let $v$ be the vector
$$
(|00\rangle+|11\rangle)
\tensor
(|00\rangle+|11\rangle)
\tensor
(|00\rangle+|11\rangle);
$$
the new code will be the span of 16 translates of $v$ under the extraspecial
group (a ``coset code'' (\cite\cosets)).

Consider the following nonlinear code of length 3 and minimum distance 2
over $GF(4)$:
$$
\align
\{
&000,011,0\omega\omega,0\omegabar\omegabar,\\
&101,110,1\omega\omegabar,1\omegabar\omega,\\
&\omega0\omega,\omega1\omegabar,\omega\omega1,\omega\omegabar0,\\
&\omegabar0\omegabar,\omegabar1\omega,\omegabar\omega0,\omegabar\omegabar1
\}.
\endalign
$$
To this code, we can associate a set of operators from the extraspecial
group, by mapping $0$ to the identity, $1$ to $Y_1$, $\omega$ to $X_1$,
and $\omegabar$ to $Z_1$.  This, then, determines a set of translates
of $v$, the span of which is easily verified to be a $((6,16,2))$.
It remains only to show that this code is nonadditive; this can be done,
for instance, by checking that some quartic invariant differs from
that of the additive $[[6,4,2]]$.
\qed\enddemo

Thus, only length 4 remains open:

\proclaim{Theorem 8}
Any $((4,4,2))$ is locally equivalent to the $[[4,2,2]]$.
\endproclaim

\demo{Proof}
Let $Q$ be a $((4,4,2))$.  We proceed by first showing that for any pair of
qubits of $Q$, there exists an orthonormal basis of the tensor product of
those qubits such that the corresponding projection operators all commute
with $P_Q$; moreover, it will turn out that, up to equivalence, the basis
can be taken to be the cosets of the $[[2,0,2]]$.  We will then consider
the consistency conditions between the bases corresponding to different
pairs of qubits; the result will follow.

From the proof of theorem 6 of \cite\invariants, we know that for any
$S\subset\{1,2,3,4\}$ with $|S|=2$,
$$
E_{v\in Q}(\Tr(-[\Tr_{S^c}(v v^\dagger)\tensor I,P_Q]^2))=0.
$$
In other words, for all $v\in Q$, $T_S(v)\tensor I$ commutes with $P_Q$,
where we define
$$
T_S(v)=\Tr_{S^c}(v v^\dagger).
$$
Consequently, for each eigenspace of $T_S(v)$, the
corresponding projection operator commutes with $P_Q$.  If $T_S(v)$ has
four distinct eigenvalues, we obtain a basis as desired.  Otherwise, let
$\Pi'$ be the projection onto an eigenspace of $T_S(v)$.  The operator
$\Pi'\tensor I$ commutes with $P_Q$, so the operator
$$
P_Q (\Pi'\tensor I)
$$
is the projection onto a subcode $Q'$ of $Q$.  Take $v'\in Q'$, and
consider $T_S(v')=\Pi' T_S(v')$.  If this separates $\Pi'$, we get the
desired basis by induction.  Suppose, on the other hand, that for all
$v'\in Q'$, $T_S(v')\propto \Pi'$.  But then the code $Q'$ can correct for
the erasure of both qubits in $S$.  This is impossible, by the Singleton
bound (\cite\knilllaflamme), unless $\dim(Q')=1$.  But then, adding up the
dimensions of the induced partition of $Q$, we conclude that we must have
partitioned $Q$ into four subspaces, giving the desired basis.

Let $v_1$, $v_2$, $v_3$, and $v_4$ be the codewords of $Q$ corresponding
to the orthonormal basis of ${\Bbb C}^2\tensor {\Bbb C}^2$ we have just
constructed.  Note that the projection onto each basis element is of the
form
$$
\Tr_{S^c}(v_i v_i^\dagger),
$$
for $i$ ranging from 1 to 4.  But then, if we trace away either remaining
qubit, we get $I/2$; it follows that each basis element is a
$((2,1,2))$.  Then lemma 6 applies; consequently, we may assume
without loss of generality that $P_Q$ commutes with the group
$$
G_{12}=\langle \sigma_x\tensor\sigma_x\tensor 1\tensor 1,
\sigma_z\tensor\sigma_z\tensor 1\tensor 1\rangle,
$$
by using an equivalence to map the orthonormal basis of ${\Bbb C}^2\tensor
{\Bbb C}^2$ associated with $S=\{1,2\}$ to the cosets of the $[[2,0,2]]$.

Consider the basis associated to $S=\{1,3\}$.  We are still free to 
transform the third qubit; this allows us to assert that $P_Q$ commutes
with
$$
T_x = (a_{xx}\sigma_x+a_{yx}\sigma_y+a_{zx}\sigma_z)\tensor 1\tensor \sigma_x
\tensor 1,
$$
with $|a_{xx}|^2+|a_{yx}|^2+|a_{zx}|^2=1$, and similarly for $T_y$ and
$T_z$.  On the other hand, $P_Q$ can be written as a linear combination of
elements of the form
$$
e_1\tensor e_1\tensor e_2\tensor e_3,
$$
where each $e_i$ is in $\{1,\sigma_x,\sigma_y,\sigma_z\}$; this follows
from the assumption that $P_Q$ commutes with $G_{12}$.  Suppose $a_{xx}\ne
0$.  Then, consider the anticommutator
$$
\{T_x,\sigma_x\tensor\sigma_x\tensor 1\tensor 1\}.
$$
On the one hand, this commutes with $P_Q$; on the other hand, it equals
$$
a_{xx} (1\tensor \sigma_x\tensor \sigma_x\tensor 1);
$$
thus
$$
1\tensor \sigma_x\tensor \sigma_x\tensor 1
$$
commutes with $P_Q$.  Suppose $a_{yx}$ were also nonzero; then
$$
1\tensor \sigma_y\tensor\sigma_x\tensor 1
$$
would also commute with $P_Q$, implying that
$$
1\tensor \sigma_z\tensor 1 \tensor 1
$$
commuted with $P_Q$, contradicting the fact that $Q$ is pure of minimum
distance 2.  So at most one of $a_{xx}$, $a_{yx}$, and $a_{zx}$ is
nonzero; since their norms add to 1, exactly one of them must be nonzero.
We can conclude, therefore, that the coefficients $a_{??}$ determine a
permutation of $\{x,y,z\}$; using the remaining freedom in the third qubit,
we may map this permutation to the identity.

Thus we can arrange for
$Q$ to commute with
$$
G_{13}=\langle \sigma_x\tensor 1\tensor \sigma_x\tensor 1,
\sigma_z\tensor 1\tensor\sigma_z\tensor 1\rangle,
$$
via transformations in the third qubit only; similarly, $G_{14}$
can be assumed to commute with $Q$.  But then $P_Q$ must be a linear
combination of
$1\tensor 1\tensor 1\tensor 1$,
$\sigma_x\tensor \sigma_x\tensor \sigma_x\tensor \sigma_x$,
$\sigma_y\tensor \sigma_y\tensor \sigma_y\tensor \sigma_y$,
and
$\sigma_z\tensor \sigma_z\tensor \sigma_z\tensor \sigma_z$.  This can
only happen when $Q$ is a coset of the $[[4,2,2]]$, and is thus equivalent
to the $[[4,2,2]]$.
\qed\enddemo

Tracing away any two qubits of a $((6,1,4))$ gives a $((4,4,2))$; this allows
us to prove:

\proclaim{Corollary 9}
Any (pure) $((6,1,4))$ is locally equivalent to quantum hexacode; that is,
the $[[6,0,4]]$ with basis
$$
\vbox{\halign{&$#$\cr
0      & 0      & 1      & 1      & 1      & 1      \cr
0      & 0      & \omega & \omega & \omega & \omega \cr
1      & 1      & 1      & 1      & 0      & 0      \cr
\omega & \omega & \omega & \omega & 0      & 0      \cr
0      & 1      & 0      & 1      & \omega & \omegabar \cr
0      & \omega & 0      & \omega & \omegabar & 1      \cr
}}
$$
\endproclaim

\demo{Proof}
Let $v$ be a (pure) $((6,1,4))$.  Then for each pair $S$ of qubits,
$$
4\Tr_S(v v^\dagger)
$$
is the projection operator onto a $((4,4,2))$, and is thus equivalent to
the $[[4,2,2]]$.  In particular, we may assume without loss of generality
that
$$
4 \Tr_{12}(v v^\dagger)
$$
is given by the $[[4,2,2]]$.  In particular, it follows that $v v^\dagger$
commutes with
$$
\align
\langle
&1\tensor 1\tensor \sigma_x\tensor\sigma_x\tensor\sigma_x\tensor\sigma_x,\\
&1\tensor 1\tensor \sigma_z\tensor\sigma_z\tensor\sigma_z\tensor\sigma_z
\rangle.
\endalign
$$
Then $4\Tr_{56}(v v^\dagger)$ commutes with
$$
\langle
1\tensor 1\tensor \sigma_x\tensor\sigma_x,
1\tensor 1\tensor \sigma_z\tensor\sigma_z
\rangle.
$$
By the proof of theorem 3, it follows that we may, by transforming only the
first two qubits, assume that $4\Tr_{56}(v v^\dagger)$ is the $[[4,2,2]]$
as well.  Thus $vv^\dagger$ must commute with the group
$$
\align
\langle
&1\tensor 1\tensor \sigma_x\tensor\sigma_x\tensor\sigma_x\tensor\sigma_x,\\
&1\tensor 1\tensor \sigma_z\tensor\sigma_z\tensor\sigma_z\tensor\sigma_z,\\
&\sigma_x\tensor\sigma_x\tensor 1\tensor 1\tensor \sigma_x\tensor\sigma_x,\\
&\sigma_z\tensor\sigma_z\tensor 1\tensor 1\tensor \sigma_z\tensor\sigma_z
\rangle.
\endalign
$$

Now, consider $P_{13}=4\Tr_{13}(v v^\dagger)$.  This must commute with
$$
\align
\langle
&1\tensor \sigma_x\tensor\sigma_x\tensor\sigma_x,\\
&1\tensor \sigma_z\tensor\sigma_z\tensor\sigma_z,\\
&\sigma_x\tensor 1\tensor \sigma_x\tensor\sigma_x,\\
&\sigma_z\tensor 1\tensor \sigma_z\tensor\sigma_z
\rangle.
\endalign
$$
It follows that $P_{13}$ can be written as
$$
\align
&\phantom{{}+{}} a \cdot{}1\tensor 1\tensor 1\tensor 1\\
&{}+b \cdot{}\sigma_x \tensor\sigma_x \tensor\sigma_y \tensor\sigma_z\\
&{}+c \cdot{}\sigma_y \tensor\sigma_y \tensor\sigma_z \tensor\sigma_x\\
&{}+d \cdot{}\sigma_z \tensor\sigma_z \tensor\sigma_x \tensor\sigma_y\\
&{}+e \cdot{}\sigma_x \tensor\sigma_x \tensor\sigma_z \tensor\sigma_y\\
&{}+f \cdot{}\sigma_y \tensor\sigma_y \tensor\sigma_x \tensor\sigma_z\\
&{}+g \cdot{}\sigma_z \tensor\sigma_z \tensor\sigma_y \tensor\sigma_x;
\endalign
$$
since $\Tr(P_{13})=4$, we must have $a={1\over 4}$.
Consider the equation $P_{13}^2=P_{13}$.  For this to hold, we must,
in particular, have that the only terms appearing in the expansion of
$P_{13}^2$ are the terms appearing in the expansion of $P_{13}$.  In
particular, consider the term
$$
1\tensor 1\tensor \sigma_x\tensor \sigma_x
$$
On the one hand, this must have coefficient 0; on the other hand, we see
that it has coefficient $2be$.  Thus $be=0$; without loss of generality,
we may assume that $e=0$.  It follows that $|b|={1\over 4}$, and
$$
\sigma_x\tensor\sigma_x\tensor\sigma_y\tensor\sigma_z
$$
stabilizes $P_{13}$.  This implies $f=g=0$ and $|c|=|d|={1\over 4}$.

But then $v v^\dagger$ commutes with the group
$$
\align
\langle
&1\tensor 1\tensor \sigma_x\tensor\sigma_x\tensor\sigma_x\tensor\sigma_x,\\
&1\tensor 1\tensor \sigma_z\tensor\sigma_z\tensor\sigma_z\tensor\sigma_z,\\
&\sigma_x\tensor\sigma_x\tensor 1\tensor 1\tensor \sigma_x\tensor\sigma_x,\\
&\sigma_z\tensor\sigma_z\tensor 1\tensor 1\tensor \sigma_z\tensor\sigma_z,\\
&1\tensor \sigma_x\tensor 1\tensor\sigma_x\tensor\sigma_y\tensor\sigma_z,\\
&1\tensor \sigma_z\tensor 1\tensor\sigma_z \tensor\sigma_x \tensor\sigma_y
\rangle.
\endalign
$$
It follows immediately that $v$ is equivalent to hexacode.
\qed\enddemo

\proclaim{Corollary 10}
Any $((5,2,3))$ is locally equivalent to length 5 Hamming code; that is,
the $[[5,1,3]]$ with basis
$$
\vbox{\halign{&$#$\cr
0      & 1      & 1      & 1      & 1      \cr
0      & \omega & \omega & \omega & \omega \cr
1      & 0      & 1      & \omega & \omegabar \cr
\omega & 0      & \omega & \omegabar & 1      \cr
}}
$$
\endproclaim

\demo{Proof}
Let $Q$ be the given $((5,2,3))$.  There exists a self-dual code $Q'$ of
length 6 such that $2\Tr_{1}(P_{Q'})=P_Q$.  But then we can compute the
weight enumerator of $Q'$; this tells us that $Q'$ is a $((6,1,4))$, and
thus locally equivalent to hexacode.  But then the code obtained by
tracing away the first qubit of hexacode must be locally equivalent to $Q$; in
other words, $Q$ is locally equivalent to the $[[5,1,3]]$.
\qed\enddemo

The proof of theorem 8 can be used to show that any $((2m,4^{m-1},2))$ with
the same quartic invariants as the $[[2m,2m-2,2]]$ is locally equivalent to
the $[[2m,2m-2,2]]$.  This suggests that it should be possible to give some
set of conditions on quartic invariants, satisfied by all additive codes,
such that any code satisfying the conditions is equivalent to an additive
code.  Note that this cannot be true of cubic invariants, since the codes
of theorem 7 have the same cubic invariants as the additive codes with the
same parameters (lemma 5 of \cite\invariants).

\head Automorphisms and equivalences\endhead

In \cite\crssII, the automorphism group of an additive code is defined to
be the subgroup of the natural semidirect product of $S_n$ and $S_3^n$ that
preserves the code.  There is also a natural concept of automorphism group
for general codes, to wit, the group of all equivalences from the code to
itself; we will call this the {\it full} automorphism group.  The full
automorphism group is of particular interest because automorphisms of a
code induce fault-tolerant operations on the encoded state
(\cite\gottesman).  It is natural, therefore, to wonder how these two
concepts of automorphism groups are related.

\proclaim{Theorem 11}
Let $Q$ be an additive code, with full automorphism group $A$, and
with automorphism group $A_0$ as an additive code.  Then the intersection
of $A$ with the Clifford group is isomorphic to the semidirect product
of $A_0$ and the centralizer of $Q$ in the extraspecial group.
\endproclaim

\demo{Proof}
We can map $S_3$ into $SO(3)$ by:
$$
\align
(12)&\mapsto \pmatrix 0&-1&0\\-1&0&0\\0&0&-1\endpmatrix,\\
(23)&\mapsto \pmatrix -1&0&0\\0&0&-1\\0&-1&0\endpmatrix.
\endalign
$$
This induces, in the usual way, a map from $S_3$ to $PSU(2)$, and
thus a map from $A_0$ to $PSU(2)^{\tensor n}$.  Let $\phi$ be
any equivalence in the image of this map.  We readily verify
that the centralizer of $\phi(Q)$ in the extraspecial group $E$
is isomorphic to the centralizer of $Q$ in $E$;
consequently, there exists some element $e\in E$ such that
$e \phi(Q)=Q$.  On the other hand, if we reduce the intersection of $A$
with the Clifford group modulo $E$, then we must get $A_0$; the result
follows immediately.
\qed\enddemo

The natural question is then: When are these groups isomorphic?  Consideration
of the $[[2,0,2]]$ reveals that it is possible for an additive code to have
nonadditive automorphisms:

\proclaim{Lemma 12}
The full automorphism group of the $((2,1,2))$ is isomorphic to
the semidirect product of $Z_2$ and $PSU(2)$, with $Z_2$ acting
as complex conjugation.
\endproclaim

\demo{Proof}
By lemma 5, we may assume that the code is $|00\rangle+|11\rangle$.
As in lemma 5, we may reduce the problem to one of unitary matrices;
we immediately find that any local equivalence must be of the form
$U\tensor \overline{U}$, for $U\in PSU(2)$.  The equivalence which
exchanges the two qubits completes the group.
\qed\enddemo

On the other hand, the larger $[[2m,2m-2,2]]$s behave much better:

\proclaim{Theorem 13}
Let $Q$ be a $[[2m,2m-2,2]]$ for $m\ge 2$.  Then every automorphism of
$Q$ lies in the Clifford group.
\endproclaim

\demo{Proof}
Since the additive automorphism group of $Q$ acts transitively on the
qubits, it suffices to consider local automorphisms.
In particular, any local equivalence corresponds to an $2m$-tuple of
elements of $SO(3)$; we need to show that every element of the $2m$-tuple
is a monomial matrix.  Without loss of generality, we may assume
that $Q$ is given by the projection operator
$$
P_Q = {1\over 4}(1^{\tensor 2m}+\sigma_x^{\tensor 2m}+
(-1)^m \sigma_y^{\tensor 2m}+
\sigma_z^{\tensor 2m}).
$$
The key observation is to note that 
$$
4 P_Q-1^{\tensor 2m}
$$
is naturally associated to the following vector in $({\Bbb R}^3)^{\tensor 2m}$:
$$
v=|000\ldots\rangle+(-1)^m|111\ldots\rangle+|222\ldots\rangle,
$$
acted on by $SO(3)^{\tensor 2m}$ in the natural way.  So the question is
then: for which elements of $SO(3)^{\tensor 2m}$ is $v$ a fixed point?

Consider the operator
$$
\langle 0|_1 \Tr_{\{3,4\ldots 2m\}}(v v^t)|0\rangle_1,
$$
acting on the second ``trit'' (that is, copy of ${\Bbb R}^3$).  We readily
see that this is proportional to $|0\rangle\langle 0|$; in particular, it
has rank 1.  Consequently, if $\phi\in SO(3)^{\tensor 2m}$ admits $v$ as a
fixed point, then the operator
$$
\langle 0|_1 \Tr_{\{3,4\ldots 2m\}}(\phi(v) \phi(v)^t)|0\rangle_1
$$
must rank 1.  Clearly, this depends only on the action of $\phi$ on the
first trit.  Thus, the condition must still be satisfied if we replace
$\phi$ by
$$
\phi'=\phi^{(1)}\tensor 1^{\tensor 2m-1}.
$$
Then we can readily compute
$$
\langle 0|_1 \Tr_{\{3,4\ldots 2m\}}(\phi'(v) \phi'(v)^t)|0\rangle_1,
$$
by first noting that
$$
\align
\Tr_{\{3,4,\ldots 2m\}}(\phi'(v) \phi'(v)^t)
&=
(\phi^{(1)}\tensor 1)
\Tr_{\{3,4,\ldots 2m\}}(v v^t)
(\phi^{(1)}\tensor 1)^t\\
&\propto
(\phi^{(1)}\tensor 1)
(
|00\rangle\langle 00|+
|11\rangle\langle 11|+
|22\rangle\langle 22|)
(\phi^{(1)}\tensor 1)^t,
\endalign
$$
since $m\ge 2$.  Selecting out the submatrix in which the first trit
is 0, we get:
$$
\pmatrix
(\phi^{(1)}_{00})^2&0&0\\
0&(\phi^{(1)}_{10})^2&0\\
0&0&(\phi^{(1)}_{20})^2
\endpmatrix,
$$
which has rank 1 if and only if exactly one of $\phi^{(1)}_{00}$,
$\phi^{(1)}_{10}$, or $\phi^{(1)}_{20}$ is nonzero.  But this must then be
true for the other rows of $\phi^{(1)}$; it follows that $\phi^{(1)}$ is a
monomial matrix.  The theorem follows immediately.
\qed\enddemo

\proclaim{Corollary 14}
Any equivalence of $GF(4)$-linear quantum codes lies in the Clifford
group, unless the codes have minimum distance 1, or contain a codeword
of weight 2.
\endproclaim

\demo{Proof}
It suffices to prove the result for local equivalence, since permuting
the qubits of a $GF(4)$-linear code gives another $GF(4)$-linear code.

Let $C$ be a w.s.d. $GF(4)$-linear code. A subset $S$ of
$\{1,2,3\ldots n\}$ will be called a minimal support of $C$ if there
exists a codeword in $C$ of support $S$, but no nontrivial codewords
of support strictly contained in $S$.  Similarly, we say a codeword of
$C$ is minimal if its support is minimal.

\proclaim{Lemma 15}
A $GF(4)$-linear code $C$ is spanned by its minimal codewords.
\endproclaim

\demo{Proof}
Let $v$ be a codeword with support $S$; we need to show that $v$ is a
linear combination of minimal codewords.  Either $S$ is minimal, and
we are done, or there exists a nontrivial codeword $v_0$ with support
$S_0$ contained in $S$.  Let $e$ be any column in $S_0$, and consider
$v_1=v-((v)_e/(v_0)_e) v_0$.  The support of $v_1$ does not contain $e$,
but is still contained in $S$.  The lemma follows by induction.
\qed\enddemo

For the associated quantum code $Q$, we first note that the minimal
supports of $Q$ can be determined from the local weight enumerator of
$Q$; consequently, the set of minimal supports is a local invariant.
Moreover, if we associate a new code $Q_S$ to each minimal support,
by selecting out those codewords of $C$ with support $S$, the code
$Q_S$ can be determined without reference to the additive structure
of $Q$.  Finally, we note that each $Q_S$ is a $[[2m,2m-2,2]]$ for
some $m$.

In particular, then, if $Q$ and $Q'$ are equivalent additive codes, then
they must have the same minimal supports, and for each minimal support, the
equivalence must take $Q_S$ to $Q'_S$.  From the hypotheses and linearity,
it follows that every minimal support has size $2m$ for $m\ge 2$.  From the
lemma, we know that $Q$ is the intersection of all the $Q_S$'s; it follows
that every qubit is covered by some minimal support.  But then it follows
from theorem 13 that the equivalence lies in the Clifford group.
\qed\enddemo

Remark.  The hypotheses of the corollary are quite weak; if $Q$ has minimum
distance 1, then some qubit of $Q$ is completely unencoded, and can thus be
removed, while if $Q$ has a codeword of weight 2, then $Q$ is the tensor
product of a smaller code and a $((2,1,2))$.  In both cases, the extra
freedom afforded is easy to determine.  It should be possible to extend
this result to additive codes, under similarly weak hypotheses.

\proclaim{Corollary 16}
If $Q$ is a $GF(4)$-linear code, then every automorphism of $Q$ lies
in the Clifford group.
\endproclaim

\demo{Proof} This follows immediately from corollary 15.\qed\enddemo

In some cases, we can apply theorem 13 to nonlinear codes:

\proclaim{Corollary 17}
Any automorphism of the $[[12,0,6]]$ is contained in the Clifford group.
\endproclaim

\demo{Proof}
The proof of corollary 16 holds, except that we consider only those
minimal supports such that three codewords have that support;
there are 6 such supports that together cover the 12 qubits.
\qed\enddemo

\proclaim{Corollary 18}
Let $Q$ be the $((5,6,2))$ defined in \cite\nonadditive.  Then
every automorphism of $Q$ is contained in the Clifford group;
in particular, its full automorphism group is the group of order 3840
given in \cite\nonadditive.
\endproclaim

\demo{Proof}
Recall from the remark following theorem 3 that, for any $((5,6,2))$,
$$
1-2\Tr_{\{1\}}(P_Q)
$$
is the projection operator of a $((4,4,2))$; in the case of the given
$((5,6,2))$, the five $((4,4,2))$s are all explicitly additive.
The result follows easily from theorem 13.
\qed\enddemo

\head Acknowledgements\endhead

The author would like to thank Rob Calderbank, Peter Shor, and Neil Sloane
for many helpful conversations.

\enddocument
\bye